# Non-Magnetic Half-Metals

Zhonghao Liu[1,2,*], S. Thirupathaiah[1,3,*], Alexander N. Yaresko[4], Satya Kushwaha[5], Quinn Gibson[5], Robert J. Cava[5] & Sergey V. Borisenko[1]

[1]*Institute for Solid State Research, IFW Dresden, D-01171 Dresden, Germany.*

[2]*State Key Laboratory of Functional Materials for Informatics, SIMIT, Chinese Academy of Sciences, Shanghai 200050, China.*

[3]*Solid State and Structural Chemistry Unit, Indian Institute of Science, Bangalore, Karnataka, 560012, India.*

[4]*Max-Planck-institute for Solid State Research, Heisenbergstrasse 1, 70569 Stuttgart, Germany.*

[5]*Department of Chemistry, Princeton University, Princeton, New Jersey 08544, USA.*

*\* These authors contributed equally to this work.*

**Half-metals are a class of materials that are metallic only for one spin direction, and are essential for spintronics applications where one needs to read, write, store and transfer spin-data[1-6]. This spin sensitivity appears to restrict them to be magnetic, and the known examples indeed are[7-13]. The fabrication of real spintronic devices from such materials is often hampered, however, by stray magnetic fields, domain walls, short spin coherence times, scattering on magnetic atoms or magnetically active interfaces, and other characteristics that come along with the magnetism. The surfaces of topological insulators[14], or Dirac or Weyl semimetals[15, 16], could be an alternative[17, 18], but production of high-quality thin films without the presence of the bulk states at the Fermi level remains very challenging[19, 20]. Here we introduce non-magnetic half-metals and demonstrate that this state is realized in IrBiSe. Using angle-resolved photoemission spectroscopy and band structure calculations we find a record-high Dresselhaus spin-orbit splitting, fully spin-polarized remnant Fermi surfaces and a chiral 3D spin-texture, all with no magnetism present. Promising applications include using IrBiSe as a source of spin-polarized electrons, and**

**lightly doped IrBiSe is expected to generate electric-field-controlled spin-polarized currents, free from back scattering, and could host triplet superconductivity.**

Materials where electrons have different energies and momenta depending on their spin orientation have significant application potential in spintronic devices[4, 21]. Manipulation of the spin degrees of freedom is best when such spin polarization and spin splitting of the electron energy bands is very strong. This strong spin differentiation is usually detected in half-metals[7–13] and on the surfaces of semiconductors[22–24], metals[25–27], metal-alloys[28] and topological insulators and semimetals[14, 16]. The former are magnetic, and the spins are usually oriented along a single global axis, thus limiting possible applications. The latter have more flexible spin textures, but are essentially two-dimensional with crystal surfaces highly sensitive to the environment, often tending to degrade rapidly over the time. Recently, several compounds have been reported that show giant spin splitting of the bulk bands, however such materials either suffer from the possibility of surface-bulk scattering since their band structures include both bulk and surface states in the vicinity of the Fermi level[29–31], or very small momentum separation of the states with different spin orientation[32]. Thus, novel materials without these drawbacks would be advantageous for addressing fundamental questions of physics and modern spintronics.

Here we present non-magnetic IrBiSe, a cubic material that has a unique electronic structure, featuring a giant (∼0.3 eV) Dresselhaus[33] spin-orbit splitting of the bulk bands near the chemical potential with an enormous momentum offset of 0.44 Å$^{-1}$ and peculiar chiral 3D spin texture.

IrBiSe is a noncentrosymmetric semiconductor, crystallizing in a simple cubic structure with a P2$_1$3 space group, which is a derivative of the structure of fools gold, FeS$_2$ (*i.e.* it is not a half Heusler compound). Its unit cell and corresponding Brillouin zone (BZ) are shown in Figs. 1(a) and 1(b), respectively. Note, that *Γ-M* is not a symmetry line. Only *Γ-X* (C2) and *Γ-R* (C3) are. The crystal structure can also be thought of as a network of dis-torted IrBi$_3$Se$_3$ octahedra with each Bi or Se ion belonging to three such octahedra (see SI).

Results of the first-principle band structure calculations of IrBiSe with and without the inclusion of spin-orbit coupling (SOC) in the computational scheme are shown in Fig. 1(c). From these calculations we find that the material is a semiconductor with a top of the valence band formed by strongly hybiridized Bi *p* - Se *p* states while the bottom of conduction band is dominated by Ir *d* $e_g$ states hybridized with the former (see SI). The bands in the vicinity of the chemical potential are strongly influenced by the spin-orbit interaction, and are spin-split everywhere in momentum space, except at the high symmetry points. Interestingly, the splitting is more significant for the valence band than it is for the conduction band. We also mention the presence of two rather rare dispersion crossings at Γ-point at -0.75 eV and -1.3 eV in the case when SOC is not included. Unlike evenly degenerate Dirac crossings two of the bands have linear dispersions while the middle one is quadratic with no linear term at Γ. They become split by SOC into two usual (-1.1 and -1.6 eV) and two double (-0.8 and -1.4 eV) Weyl points. The most remarkable feature of the electronic structure of IrBiSe is the top of the valence band formed by the strongly dispersive state along the *Γ-M* direction. As follows from the comparison of the computational results, these are non-degenerate states, well split from the other bands below the gap. There are twelve such maxima in the cubic BZ, shown in Fig. 1(b), as dictated by symmetry.

The high-quality single crystals used for the experimental study (see SI and Fig. 1(d)) were characterized by transport (Fig. 1(e)) and integrated photoemission measurements (Fig. 1(f)). The heights of the characteristic peaks in the latter spectrum confirmed the stoichiometric composition. The electrical resistance of IrBiSe as a function of temperature is shown in Fig. 1(e), from which it is clear that IrBiSe is a semiconductor as the resistance increases rapidly upon cooling the sample. Further, Fig. 1(e) demonstrates that IrBiSe is a nonmagnetic semiconductor as the electrical resistance shows no variation between zero field cooling and field cooling.

The ARPES data obtained for IrBiSe are compared with corresponding results of the calculations in Fig. 2. The photoemission signal from semiconductors is typically very low unless the surface states are present. We have covered many BZs by using different photon energies and scanning large portions of $k$-space. In all of the numerous cleaves performed have we noted the absence of surface states. Therefore in the following we attribute observed features in the spectra to the bulk spectral function. In Figs. 2(a) and 2(b) we compare the momentum distribution of the spectral weight within a narrow energy window near the chemical potential with the calculated result integrated over $k_z$. As expected from the band structure shown in Fig. 1(c), projection on the (001) surface should give four in-plane features along the diagonals of the square BZ and another four features corresponding to the other eight out-of-plane maxima (each pair of them will have the same projection). The experimental map reproduces these multiple spot-like features. We compare with integrated calculated result because of intrinsically moderate ARPES $k_z$-resolution and because at a given photon energy momentum distribution map corresponds to constantly decreasing $k_z$ values upon increasing absolute $k_{//}$. In the first BZ, centered at (0, 0) in Fig. 2(b), one sees four features on the diagonals meaning that $k_z$ here corresponds mostly to $\Gamma$-plane, whereas in other repeated BZ the spots lying on verticals and horizontals are more pronounced.

We do not expect a full quantitative correspondence between the datasets shown in panels (c) and (d) of Fig. 2 because the theoretical cut corresponds exactly to the diagonal of the BZ ($\Gamma$-$M$) and in particular $k_z=0$ while the experimental one is slightly off diagonal, clearly catching only one of the low-energy features, and is broadened by resolution, temperature and matrix element effects. Taking into account that the data are taken from the insulator, the overall agreement is very good and it is clear that the origin of the spots on the momentum distribution map is exactly the most intriguing feature of IrBiSe – namely the presence of parabolic dispersion that is clearly split from the rest of the bands. Fig. 2 further suggests that the detected experimental bands are of only bulk origin.

We further examine the parabolic-like dispersion with the maximum at the chemical potential in more detail. We have recorded the momentum-energy intensity distribution using different photon energies along the momentum cuts running parallel to the black arrow through the pair of closest spots on the map from Fig. 2(b) and zoomed in towards the chemical potential (Fig. 3(a) and 3(b)). The datasets correspond to different $k_z$-values showing dispersions making two diagonal spots (Fig. 3(a)) and two spots from one horizontal and one vertical (Fig. 3(b)). The correspondence with calculated data (right panels of Figs. 3(a) and 3(b)) is remarkable, with small inconsistencies related to the misalignment mentioned earlier. The splitting of the bands is clearly seen, allowing one to read its magnitude from the second-derivative maps shown in middle panels of Figs. 3(a) and 3(b). From the experimental data we estimate a spin-splitting of 270 meV, which is lower than the predicted value of 360 meV from the calculations. On the other hand, the experimental momentum offset of 0.4 Å$^{-1}$ is consistent with the theoretical value of 0.43 Å$^{-1}$.

We also estimate the spin character of two closest features by using circularly polarized light in Fig. 3(c). We are aware that circular dichroism cannot be directly related to the spin-polarization because of ARPES matrix elements, but in this case we evaluate the signal from the features that are not symmetric with respect to any experimental plane; the observed strong dichroism can be considered as evidence for the very different spin characters of these features.

With the help of the band structure calculations shown in Fig. 1, we estimated the spin splitting ($\Delta E$) as a function of momentum ($k$) along different high symmetry directions, as shown in Fig. 3(d), from the upper valence bands. In the figure, the curves with markers are the data, while the solid black lines are fits to the data using the equation $\Delta E = \alpha k + \gamma k^3$. The fits are good up to a momentum vector of 0.23 Å$^{-1}$ (47% of $\Gamma$-$X$ width) from the zone center as shown in Fig. 3(d). From these fits we find that the spin splitting is linearly dependent on $k$ for the bands in the $\Gamma$-$X$ direction, while it is cubically dependent on $k$ for the bands dispersing in rest of the high-symmetry directions ($\Gamma$-$M$ and $X$-$R$). Furthermore, with the help of these fits we can estimate the SOC parameters $\alpha$ and $\gamma$. In the $\Gamma$-$X$ direction we estimate splitting parameters $\alpha$=0.44 ±0.01 eV Å and $\gamma$= 0, in the $\Gamma$-$M$ direction there are $\alpha$=0 and $\gamma$=9.94 ±0.14 eV Å$^3$ and, in the $Z$-$K$ direction there are $\alpha$=0 and $\gamma$=10.74±0.1 eV Å$^3$. The values of $\gamma$ are in very good agreement with the predicted values for GaAs ($\gamma$=8.3 eV Å$^3$ [34] and 9.13 eV Å$^3$ [35]).

Finally, we present the theoretically defined spin characters of the states in IrBiSe in Fig. 3(e). Spins projections onto the directions parallel and perpendicular to the given high symmetry direction and to the $k_z$-axis are given by the size of the marker. Judging from the uneven distribution of the markers and their sizes on the plots, one can easily conclude that the electronic states near chemical potential of IrBiSe display a peculiar spin texture. The occupied states along $\Gamma$-$X$ have spins oriented parallel to this direction, since middle and right panels containing $\Gamma$-$X$ show virtually no markers. The spin-polarization of the discussed earlier highest occupied states is also well-defined. These spins are directed almost perpendicular to $\Gamma$-$M$, having no $k_z$-component, thus strongly reminiscent of the spin-momentum locking on the surfaces of the topological insulators[17].

We thus have seen that IrBiSe has a very peculiar electronic structure: there is only one, non-spin-degenerate and well split from others, dispersive electronic feature that approaches the chemical potential, with spins directed nearly perpendicular to the momentum vector. This dispersion gives rise to the intriguing remnant Fermi surface consisting of spots at discrete values of momentum. To settle the terminology, we briefly mention that the detected giant spin-orbit splitting is of the Dresselhaus type[33] since it is observed for the bulk states due to the lack of inversion symmetry. In analogy to topological insulators, we will refer to IrBiSe as a non-magnetic half-metal (NMHM) since its metallicity is a matter of light hole doping, which appears to be possible, and its Fermi surface is supported by electronic states having only one spin orientation. Unlike in ferro- or antiferromagnetic half-metals[7-13] the projection of the spin to the local axis rather than to a global axis is meant here. Obviously, such a unique and simple arrangement of valence electrons should imply the presence of interesting physics and a range of useful applications.

Already pristine IrBiSe can be used as an alternative to GaAs as a source of spin-polarized electrons (Fig. 4(a)). Indeed, the first photoelectrons coming out of the sample upon hypothetical increase of photon energy of the exciting radiation from zero will be spin-polarized and in order to flip the spin-polarization, one would just need to tilt or rotate the sample (instead of changing the light helicity, as in GaAs case[36, 37]), switching from spot to spot of the remnant Fermi surface. If one filters out only electrons with highest kinetic energies, it is possible to use practically any coherent light source capable of ejecting the photoelectrons. Moreover, as the band structure shown in Fig. 1(c) suggests, the lowest energy optical transition (0.81 eV) corresponds exactly to the spin-polarized states with the band maximum along $\Gamma$-$M$, although the gap is indirect. One can thus apply techniques, similar to the GaAs case, where cesium oxide treatment of the surface was used to decrease the electron affinity of the material[36] and use a laser with a wavelength of ~1530 nm to ensure the emission of spin-polarized electrons with minimal momentum.

The Fermi surface of lightly hole-doped IrBiSe is presented in Fig. 4(b). From the previously studied underlying band structure it is clear that the pockets are hole-like and are created by the same dispersive feature with its maximum along $\Gamma$-$M$. Though the pockets are identical, their shape is not symmetric with respect to $\Gamma$-$M$, which is dictated by the symmetry of the irreducible part of the BZ (see Fig. 1(b) and SI). If projected to any of the coordinate planes, they will correspond to the spots on the remnant Fermi surface map shown earlier in Fig. 2(a) and 2(b). Each pocket is singly degenerate, meaning that only a band with one spin character crosses the Fermi level. This is why such material can be called a half-metal. We have checked that all electronic states forming such small pockets have the same direction of spins in the local coordinate system (see SI): they are all almost perpendicular to the momentum. Such an arrangement of spins results in a unique three-dimensional spin texture shown in Fig. 4(c). An immediate association with the spin texture of the surface states in topological insulators springs to mind[14, 17], but the case discussed here for IrBiSe is a 3D generalization of the latter. The spins are not tangential to the circle, but nearly tangential to the twelve points on the sphere in momentum space. Such a spin texture of the states at the Fermi level guarantees the absence of backward scattering events in any of these directions since the spins for opposite momenta are always opposite to each other. This is a very desired property for a spintronics-relevant material where long coherence times of spin-polarized electrons are essential, as is a 3D single crystal material with a simple cubic crystal structure.

Recent experimental studies have reported electrical detection of earlier-predicted current-induced spin polarization on the surfaces of topological insulators[38, 39]. We anticipate a similar effect in lightly hole-doped IrBiSe, but that the spin-polarized currents will be generated in the bulk, rather than on the fragile surface. Indeed, application of an electric field along any of the twelve directions corresponding to $\Gamma$-$M$ (see Fig. 4(d)) will produce a net momentum along the $k_x$ direction. This is indicated by the shifted projection to the ($k_x$, $k_y$) plane of the sphere mentioned earlier; the top-view is shown in the upper part of Fig. 4(d). This shift will result in an electric current along $x$ which also induces a spin-current with spins oriented along -$y$. Thus, by putting a usual current through such a sample one can generate a flow of spin-polarized electrons and its amplitude and direction can be controlled by the initial current.

Considering the 3D spin texture along the (111) direction, one notices its chirality, dictated by the C3 symmetry of the crystal structure. Negative and positive helicities are seen in Fig. 4(e) and 4(f) respectively. This is not surprising taking into account that a center of inversion is absent in IrBiSe. This chirality should give a different response when interacting with circularly polarized light and could possibly be used as a spin sensor, *e. g.* in spin-polarized STM tips. It is a very intriguing to ask whether doped IrBiSe will become superconducting at low temperatures. On one hand, when the pockets are very small, there are ideal conditions for electrons to pair up since all of them will have a counterpart with opposite $k$ and $s$. On the other hand, when the doping is sufficiently large, so that the electrons can pair up within a given pocket, the superconductivity will be of the triplet type since all electrons corresponding to this pocket will have the same spin.

We believe that non-magnetic half-metals are a new materials class in general, and that IrBiSe in particular will play a significant role in solving both fundamental puzzles of physics and practical problems of realization of highly efficient spintronic devices.

**Acknowledgements** This work was supported under DFG grant 1912/7-1. Z. L. acknowledges support by the National Natural Science Foundation of China, Strategic Priority Research Program (B) of CAS, and Shanghai Sailing Program. S. T. acknowledges support by the Department of Science and Technology, India through the INSPIRE-Faculty program (Grant No.: IFA14 PH-86). The work at Princeton University was supported by NSF MRSEC grant DMR 1420541 and the ARO MURI on topological insulators, grant W911NF-12-1-0461. We are grateful to Martin Knupfer for the fruitful discussion.

**Competing Interests** The authors declare that they have no competing financial interests.

**Correspondence** Correspondence and requests for materials should be addressed to S. V. B. (email: S.Borisenko@ifw-dresden.de) and to R. J. C. (email: rcava@princeton.edu).

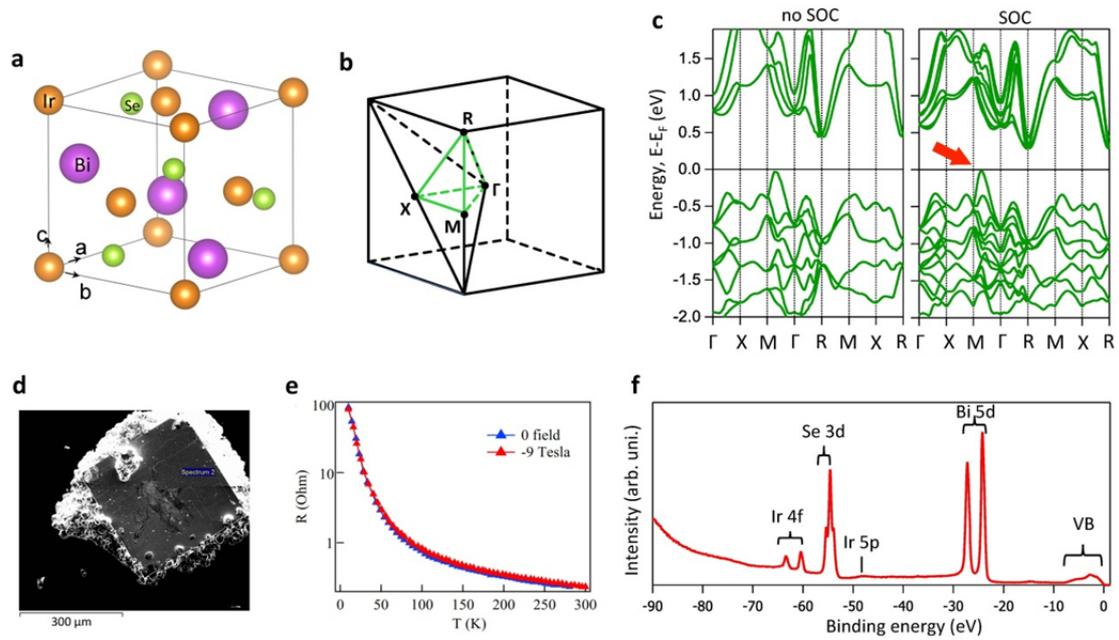

**Figure 1** (a) The crystal structure of IrBiSe. (b) The Brillouin Zone (BZ) and high symmetry points of the simple cubic structure. Note that the actual irreducible part is larger than volume enclosed be green lines due to absence of center of inversion. (c) The band structure from the first-principle calculations without and with spin-orbit coupling (SOC). (d) Single crystal of IrBiSe. (e) Temperature-dependent electrical resistance. (f) Integrated photoemission spectrum.

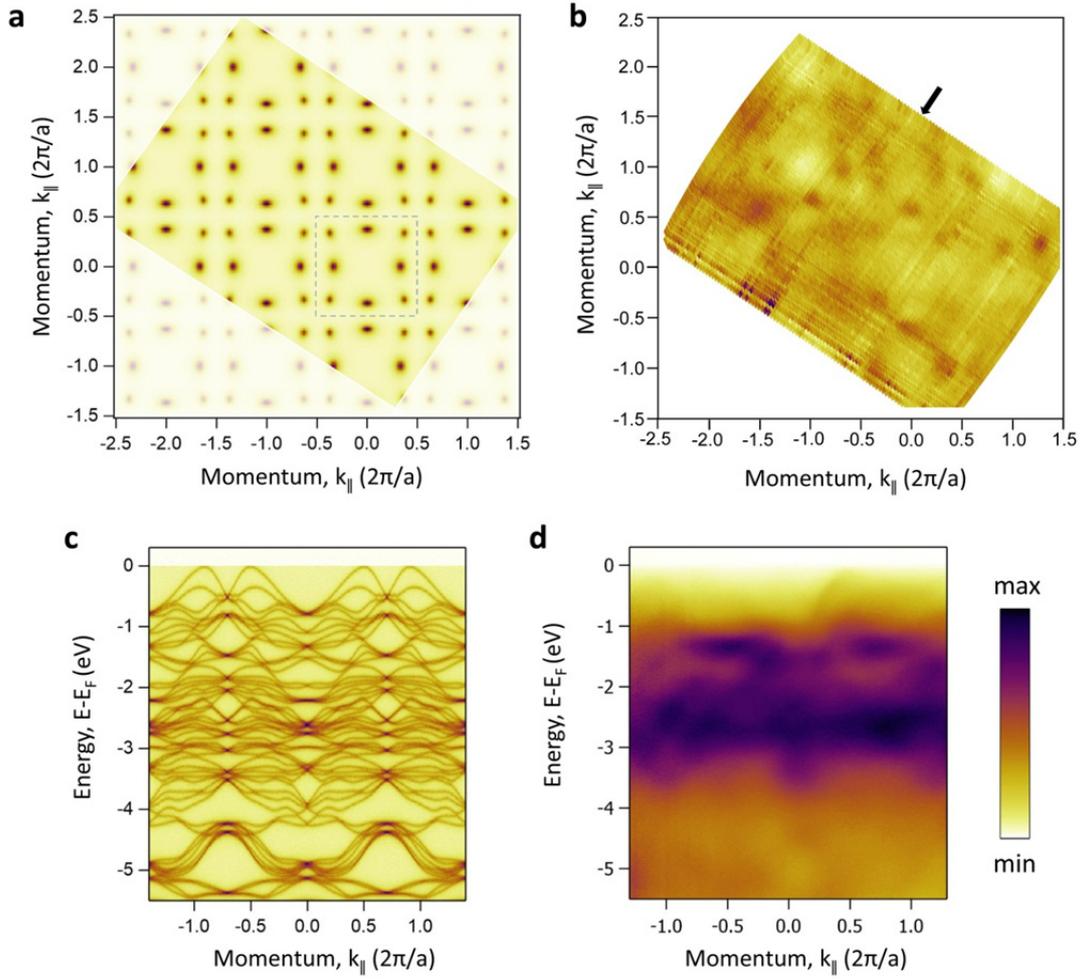

**Figure 2** (a) Calculated spectral weight just below the chemical potential and integrated over $k_z$. (b) Remnant Fermi surface map obtained by integrating over an energy window of 50 meV centered at the chemical potential using a photon energy of 85 eV. (c) Calculated spectral weight along the diagonal $\Gamma$-$M$ direction at $k_z$=0. (d) Photoemission intensity recorded along the direction indicated in the panel (b) by black arrow, $h\nu$=85 eV.

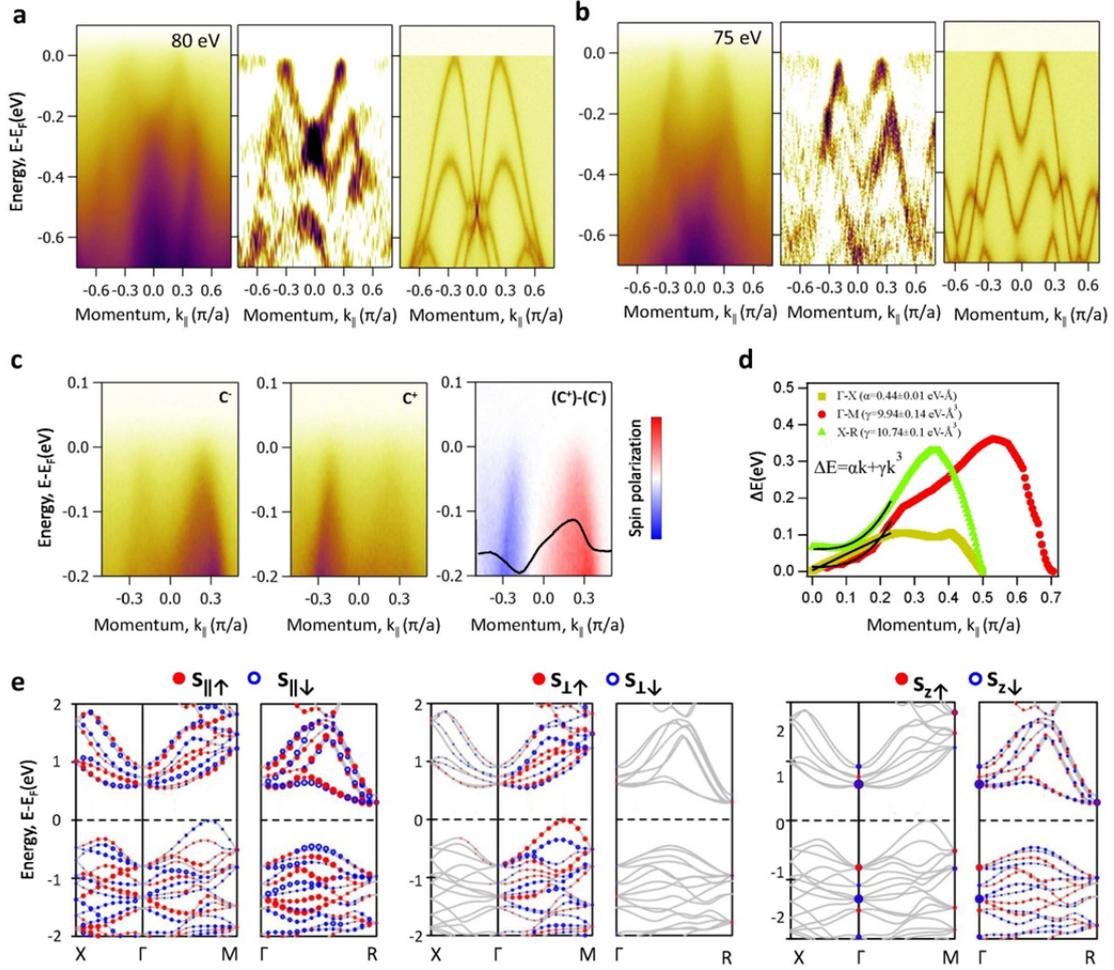

**Figure 3** (a) and (b) ARPES intensity, its second derivative plots recorded at the same angle as the arrow indicates in Fig. 2, but along the cuts going through the closest spots of the remnant Fermi surface, measured using 80 eV and 75 eV photons respectively. Right panels are corresponding calculated dispersions: along *Γ-M* centered at the M-point and along the line crossing spots form horizontals and verticals. (c) Spectra recorded along directions similar to those shown for the (a) and (b) momentum cuts with light of negative and positive helicities and their difference *i.e.,* the circular dichroism. (d) Spin splitting (*ΔE*) as function of momentum (*k*) in different high-symmetry directions starting from the zone center. The black solid curves are the fits to the data using the equation shown in the figure. (e) Spin-polarization of the bands in IrBiSe. $s_{//}$ is a spin projection along a *k* direction. $s_\perp$ and $s_z$ are spin projections perpendicular to the *k* direction. One can see that the spins are parallel to *k* along (*Γ-X*), nearly orthogonal to *k* along (*Γ-M*) and nearly parallel to k along (*Γ-R*).

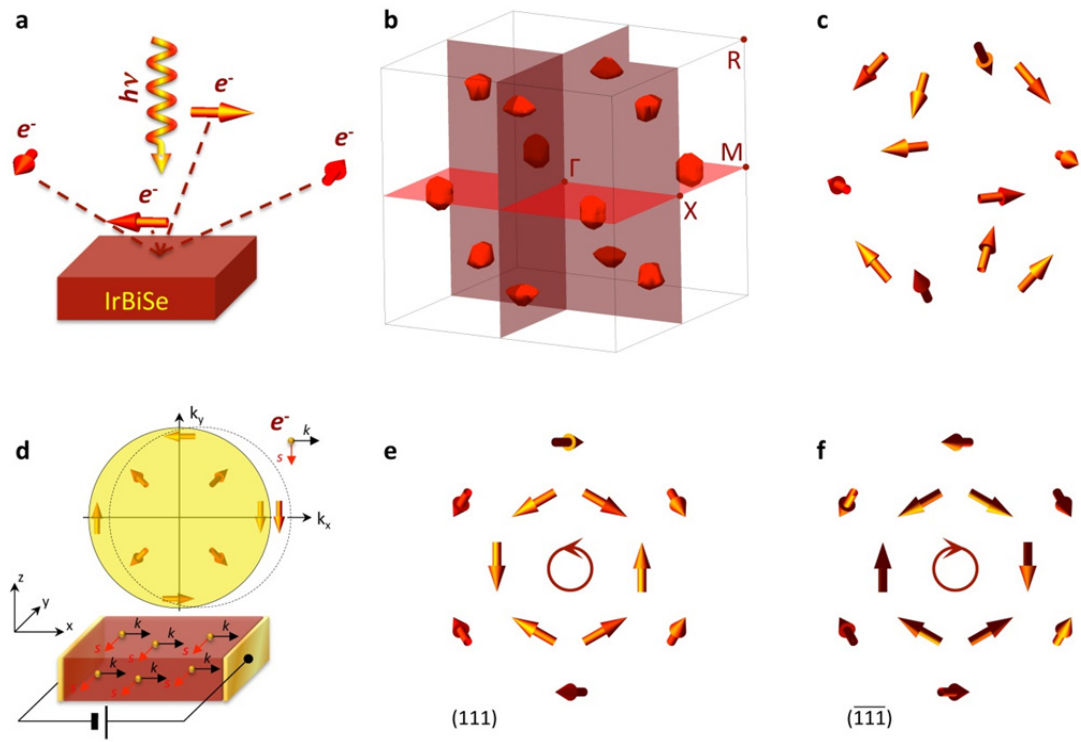

**Figure 4** (a) Schematic illustration of the emission of the spin-polarized electrons stimulated by photons. Here and in panel (d) sides of the crystal are along Γ-M. (b) Fermi surface of lightly doped IrBiSe obtained by cutting the electronic structure of pristine compound 50 meV below the chemical potential. (c) Spins corresponding to the pockets of Fermi surface from (b). Here and in the following panels, spins are considered as perpendicular to momentum vectors. (d) Schematics of generation of the spin-polarized current. (e) and (f) Views on the spin texture from (111) and ($\bar{1}\bar{1}\bar{1}$) directions respectively.

**Supplementary Information:** The single crystals of IrBiSe were grown by a self-flux method. The required amounts of the elements with high purity were sealed in a quartz tube and heated to 1000°C, held for 6 hours to allow homogeneous mixing of the elements, followed by slow cooling. The crystals were separated from the flux by centrifuging at 800°C. The well-facetted cubic crystals were successfully harvested, and an electron microscope image of a typical crystal is shown in Fig. 1. The elemental composition of crystals was evaluated by energy dispersive X-ray spectroscopy (EDS). The grown crystals possess 1:1:1 atomic composition, as evaluated by EDS, given in Table S1. More information on the crystal structure of IrBiSe can be taken from Fig. S1.

ARPES measurements were performed at the "1³-ARPES" end station. Total energy resolution was set between 5 and 10 meV, depending on the applied photon energy. Samples were cleaved *in situ* and measured at a temperature of 20 K.

Self-consistent band structure calculations were performed on the cubic crystal structure of IrBiSe having the lattice constants a=6.33 Å, using the linear muffin-tin orbital (LMTO) method in the atomic sphere approximation (ASA) as implemented in PY LMTO computer code[1]. The Perdew-Wang parameterization[2] was used to construct the exchange correlation potential in the local density approximation (LDA). Spin-orbit coupling was taken into account at the variational step. Calculations were done using fully relativistic LMTO which uses 4-component solutions of the Dirac equations.

Relativistic bands together with the orbital composition are shown in Fig.S2. It is seen that the top of the valence band is formed mostly from Bi *p* states hybridized with Se *p* and Ir *d* states. Finally, in Figs. S3 and S4 we present the results of the calculations regarding the spin-polarization of the bands in the global and local coordinate systems respectively.

**Table 1**: Elemental composition of crystals.

| Element | Weight % | Atomic % |
|---------|----------|----------|
| Se L    | 17.91    | 32.97    |
| Ir M    | 43.90    | 33.19    |
| Bi M    | 48.66    | 33.84    |

**Figure 1** Crystal structure of IrBiSe.

**Figure 2** Relativistic fat bands.

**Figure 3** Spin-polarization of the bands in global coordinate system.

**Figure 4** Spin-polarization of the bands in local coordinate system.

## structure

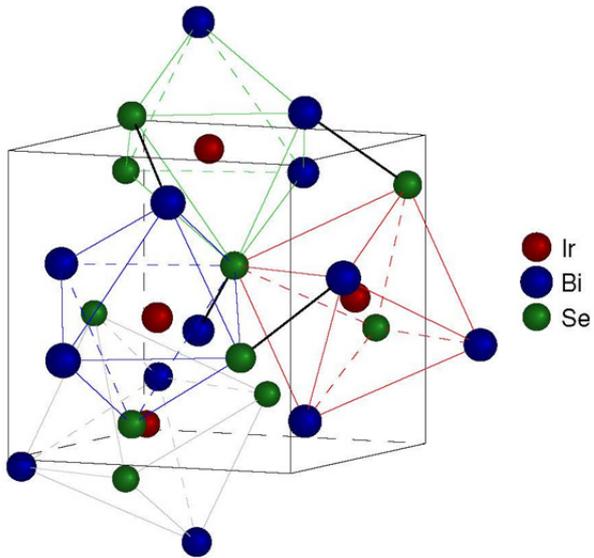

$d_{\text{IrBi}} = 3 \times 2.717, 3 \times 2.517$ Å
$d_{\text{BiSe}} = 1 \times 2.680$ Å
$d_{\text{BiSe}} = 3 \times 3.483, 3 \times 3.573$ Å
$d_{\text{BiBi}} = 6 \times 3.852$ Å
$d_{\text{SeSe}} = 6 \times 3.853$ Å

symmetry operations:
$E$, $C_{2,100}$, $C_{2,010}$, $C_{2,001}$
$C^{\pm}_{3,111}$, $C^{\pm}_{3,\bar{1}11}$, $C^{\pm}_{3,1\bar{1}1}$, $C^{\pm}_{3,1\bar{1}\bar{1}}$

- distorted IrBi$_3$Se$_3$ octahedra (anti-prisms)
- each Bi or Se ion belongs to 3 octahedra
- Bi–Se "dimers"

# relativistic fat bands

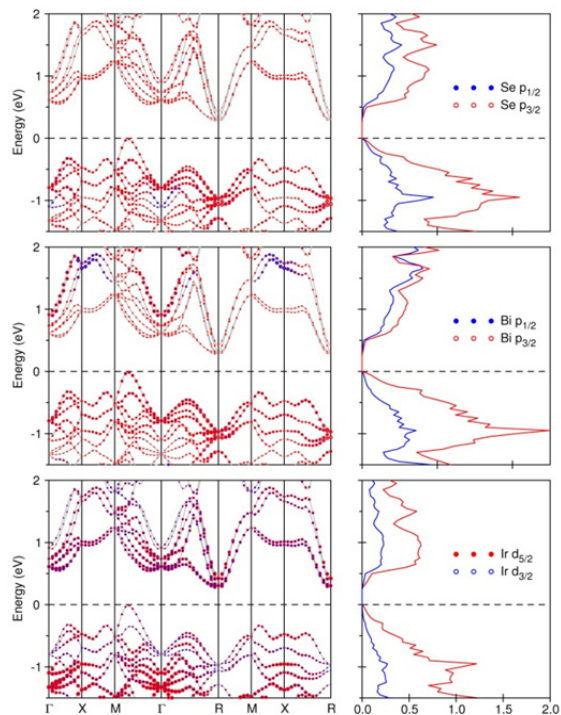

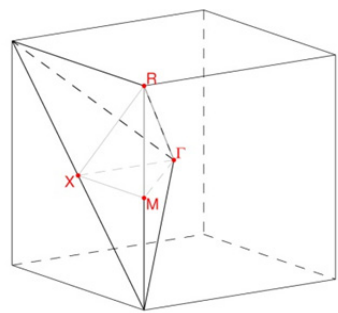

IBZ=1/12 BZ

- Bands at the top of the valence band are formed predominantly by Bi $p_{3/2}$ hybridized with Se $p$ and Ir $d$

# band spin-polarization (global frame)

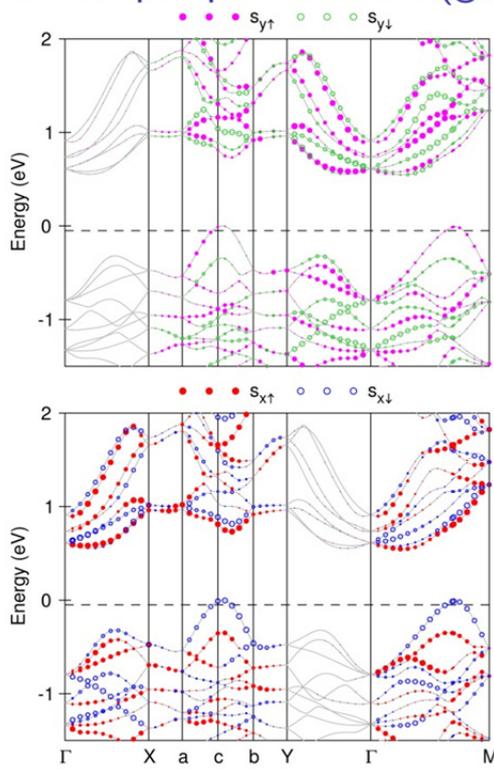

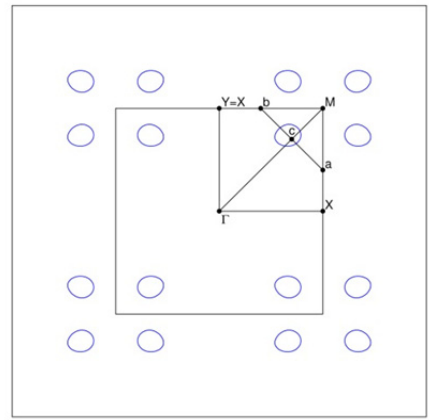

$\Delta E_F = -0.05$ eV

- strong band spin-polarization near the top of the valence band
- in the $\Gamma$XMY plane $s_z = 0$
- $s_y = 0$ ($s_x = 0$) along $\Gamma$-X ($\Gamma$-Y)

## band spin-polarization (local frame)

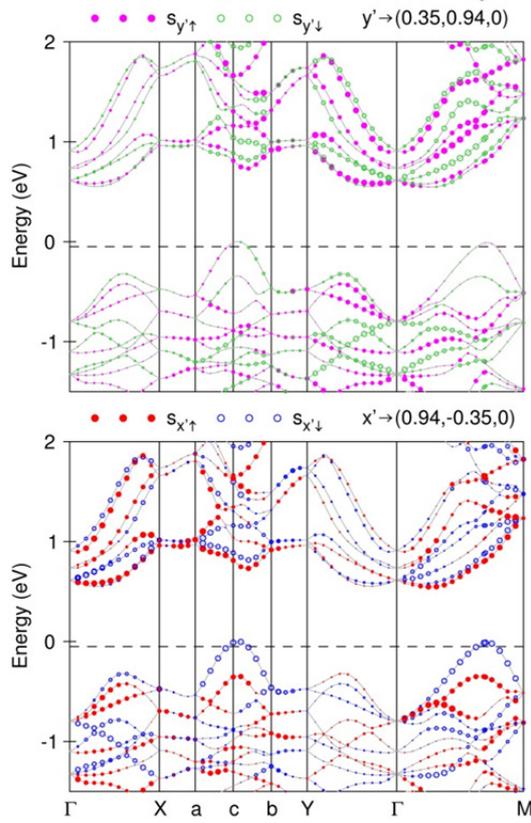

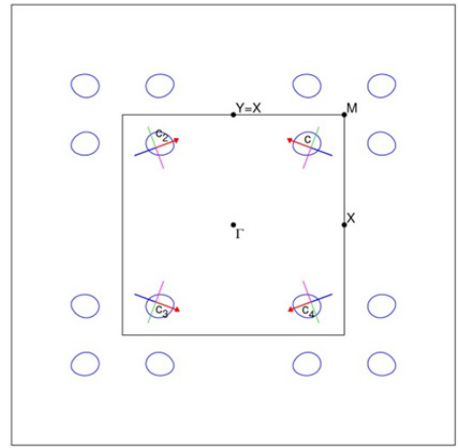

$\Delta E_F = -0.05$ eV

- for each hole pocket spin points approximately along the local $x'$ axis
- $s_{x'}$ and $s_{y'}$ are plotted for $c$ pocket
- $c_2 = C_{2,y}c$, $c_4 = C_{2,y}c$
  $c_3 = C_{2,z}c = \Theta c$; $\Theta$ is time reversal